\documentstyle[pre,aps]{revtex}
\begin{document}

\title{Dynamics of Dry Friction: A Numerical Investigation}
\author{Y. F. Lim and Kan Chen}

\address{ Department  of  Computational Science,
National University  of Singapore, Singapore 119260}

\date{\today}

\maketitle
\begin{abstract}
We perform extended numerical simulation of the dynamics of dry 
friction, based on a model derived from the phenomenological 
description proposed by T. Baumberger et al.~\cite{baumberger1,heslot}. 
In the case of small deviation from the steady sliding motion, 
the model is shown to be equivalent to the state- and 
rate-dependent friction law which was first introduced by Rice and 
Ruina \cite{ruina} on the basis of experiments on rocks. 
We obtain the dynamical phase diagram that agrees well with 
the experimental results on the paper-on-paper systems. 
In particular, the bifurcation between stick-slip and steady sliding 
are shown to change from a direct (supercritical) Hopf type to an 
inverted (subcritical) one as the driving velocity increases, 
in agreement with the experiments. 

\vspace{4pt}
\end{abstract}

\section{\bf Introduction}
It is well-known that the frictional resistance is independent 
of the apparent area of the sliding surface and it is proportional 
to the normal load with a 
proportional constant $\mu$, which is known as the friction coefficient. 
Traditionally, the friction coefficient has two distinct values: the 
static friction coefficient $\mu_{s}$, determined from the minimum force 
needed to move a slider at rest, and the dynamic friction coefficient $\mu_{d}$, used for the
 friction force when a steady sliding motion is established. 
Friction is often described using the well-known 
Amontons-Coulomb's laws: ($i$) both $\mu_{s}$ and $\mu_{d}$ are 
independent of the apparent area of the contacting surfaces and the normal 
load; ($ii$) both $\mu_{s}$ and $\mu_{d}$ depend on the shear characteristics 
of the contacting materials \cite{bowden,rabinowicz}; ($iii$) for most cases, $\mu_{d}$ is appreciably lower than $\mu_{s}$. This standard 
picture is still widely 
accepted nowadays; however, in many cases, significant refinements must be 
taken into account for the explanation of the observed friction. 

Our common experiences tell us that the sliding of contact bodies 
subjected to a steady pulling velocity sometimes proceeds in an alternation 
of periods of rest and slide rather than moving steadily (this is the
reason, for example, for the occurrence of squeaking noises). This unstable 
motion consists of periods of a stick state followed by a sudden slip is 
known as stick-slip oscillations. The stick-slip motion occurs 
on both large and small scales. Understanding stick-slip phenomenon is not
only important for many engineering applications, but also for understanding the mechanism of earthquakes (which is a 
stick-slip phenomenon at geological scale \cite{brace}).

Obviously, stick-slip is caused by the variation of frictional resistance 
during sliding. In certain range of the velocity $V$, a velocity weakening 
behavior (corresponding to a decreasing function $\mu_{d}(V)$ with respect to the steady relative sliding velocity $V$ of the contact surfaces) 
is often observed in many different materials including metals \cite{rabinowicz} and rocks \cite{scholz}. In this range, which 
generally lies in low-velocity regime, the 
steady motion is unstable with respect to perturbation; this gives rise to  
stick-slip phenomena. On the other hand, the static friction coefficient 
$\mu_{s}$ is generally found to be an increasing function with respect to 
the contact age of contact surfaces \cite{scholz,oden}. This can be explained by the fact that the plastic relaxation of the stressed 
contact junctions during a stick period leads to an increasing 
real contact area, hence strengthening the 
contact. The mechanism for the velocity-dependence of $\mu_d$ and the
mechanism for the age-dependence of $\mu_s$ are not un-related; both 
can be understood using the concept of a memory length \cite{scholz}. 

The purpose of the paper is to show, by explicit numerical
simulation, that the model based
on the phenomenological description of Ref. \cite{heslot} can be used to
explain qualitatively (and sometimes even quantitatively) 
many interesting features found in experiments. 
 The model incorporates the existing understanding 
of the age-dependence of the
static friction and the velocity-dependence of the dynamic friction
at low velocity, but does not make a distinction between the dynamic
and static frictions. There is no abrupt change from ``static''
friction to ``dynamic'' friction as in the traditional description
of friction; this is important for numerical simulation used to construct
the entire phase diagram for the dynamics of dry friction. The rest of
the paper is organized as follows. We first review the phenomenology
of dry friction, focusing on the recent experiments by T. Baumberger et
al.~\cite{baumberger1,heslot}. We then show that the 
phenomenological description of Ref.~\cite{heslot} is equivalent 
to the well-known Rice-Ruina friction law
on rock friction in the case of small deviation from the constant
velocity motion \cite{ruina}. The results from simulation of the model will
be presented in the last section together with some concluding
remarks.

\section{Phenomenology of Dry Friction}

Following the pioneering work of Rabinowicz on metals \cite{rabinowicz},
many experimental studies have been performed to study
the low velocity friction properties of various materials.
It is beyond the scope of the paper to review these
experimental work on the dynamics of dry friction. Instead,
we focus on a recent extensive experimental study of the dry friction dynamics of a paper-on-paper block-spring system (shown in Fig.~1) 
by T. Baumberger et al.~\cite{baumberger1,heslot}. Their experiments verified many known properties of dry friction and gave a 
rather complete picture of dry friction  
in various regimes. In particular, they explored systematically
the dynamical phase diagram by varying the driving velocity $V$, the slider mass $M$ and the spring stiffness $K$. The features of their
experiments are summarized as follows. 
  
\begin{itemize}
\item The phase diagram in control parameter space $(V,K/M)$ consists of two regions that can be characterized by stick-slip and steady sliding 
respectively and they are separated by a bifurcation curve. The character 
of the bifurcation changes from a direct (supercritical) Hopf type in the 
creep-dominated regime to an inverted (subcritical) one in the inertial 
regime as $V$ increases. 
    
\item In the steady sliding region, the measured dynamic friction coefficient $\mu_{d}$ exhibits velocity weakening in low-velocity range ($\leq 0.1mm/s$) which can be fitted as 
\begin{equation}
\label{mu_dV} 
\mu_{d}(V)\mid_{low V}=a_{v}-b_{v}\ln (V/V_{0}),
\end{equation} 
where $V_{0}$ is an arbitrary velocity scale; and velocity strengthening at 
larger velocities which can be characterized as 
\begin{equation}
\label{mu_d_highV}
\mu_{d}(V)\mid_{high V}=\mu_{d}^{0}+\eta V.
\end{equation}
   
\item The static friction coefficient 
$\mu_{s}$ was found to increase with the contact age $t_{st}$ as 
\begin{equation} 
\label{mu_st}
\mu_{s}(t_{st})=a_{s}+b_{s}\ln (t_{st}).
\end{equation} 

\item By introducing a characteristic memory length $D_{0}$, a relation 
between Eq.~\ref{mu_dV} and Eq.~\ref{mu_st}, i.e.
\begin{equation}
\mu_{d}(V)\mid_{low V}=\mu_{s}(D_{0}/V)
\end{equation}
can be established. Here $D_{0}$, which is given experimentally 
as $0.9\mu m$ \cite{heslot}, can be interpreted as an average 
sliding displacement needed to move to new micro-contacts. 
 
\end{itemize} 

These results are rather general (rock-rock friction, 
for example, exhibits similar features). 
The experiments show that the motion of the system at low-velocity 
($\leq 0.1mm/s$) is primarily controlled by a creep process. Based 
on these results, a phenomenological model of dry friction dynamics at 
low-velocity regime \cite{heslot} of the paper-on-paper system has been proposed. Both linear \cite{heslot} and 
nonlinear \cite{baumberger2} stability analysis of the model near the bifurcation give 
excellent quantitative agreements with experiments. Beside this, the transient behavior in the steady sliding region of the system after 
setting the driving velocity to zero suddenly has been studied under 
the aid of the model \cite{baumberger3}, where two-stage process has 
been observed. However, there is no systematic theoretical
study of the entire regime covering the crossover from the
``static'' friction regime to the ``dynamic'' friction regime and the crossover to the high-velocity (inertial) regime. In this paper we show
numerically that most important experimental features in various
regimes can be reproduced in a single model.  
 
\section{A Phenomenological Model} 

We follow the phenomenological approach used in Ref. \cite{heslot}.
Consider the experimental setup shown in Fig.~1.
The motion of the block under the influence of the external driving 
force as shown in Fig.~1, is assumed to be a thermally activated creeping 
motion in a periodic pinning potential biased by the external driving force.  In the low bias regime, where the barrier heights of the effective 
potential are comparable with the thermal activation energy, the velocity 
of the slider is given by 
\begin{equation} 
\label{dot_x}
\dot{x}=a(\frac{1}{\tau_{+}}-\frac{1}{\tau_{-}}), 
\end{equation} 
where $a$ is the typical distance between two potential minima, and $\tau_{+}$  ($\tau_{-}$) is the thermal time for escaping from a 
given well into its downstream (upstream) nearest neighbor. 
If we let the corresponding barrier height to be $U_{+}$ ($U_{-}$), 
then we have 
\begin{equation}
\label{tau_inverse}
\frac{1}{\tau_{\pm}}=\frac{\omega_{0}}{2\pi}\exp \{-\frac{\Delta U_{\pm}}
{\sigma}\},
\end{equation}  
where $\sigma$ is the thermal activation energy. $\sigma$ can be written as $N_{cr}RT$, where $N_{cr}$ is the number moles of the degrees of 
freedom involved in the creep motion. $\omega_{0}$ is the 
oscillation frequency about the minimum of the effective potential. 
The amplitude $\Delta U_{0}$ of the periodic pinning potential is 
assumed to increase with the dynamical contact age 
variable $\phi$ defined in Ref. \cite{heslot}, hence the barrier heights are given by 
\begin{equation}
\Delta U_{\pm}=\Delta U_{0}(\phi)\mp F_{ext}a/2,
\end{equation}
where $F_{ext}$ is the external force (in the experimental setup,
it is the spring force induced by the driving velocity $V$). 

Although Eqs.~\ref{dot_x} and \ref{tau_inverse} are valid strictly only 
for a time- and position-independent 
external force and a time-independent pinning potential, they can also be
used in more general cases under the assumption that 
the changes in $F_{ext}$ and $\Delta U_{0}(\phi)$ are so slow that 
an \emph{adiabatic} approximation is valid. Combining the
above equations we can write the velocity of the slider as 
\begin{equation}
\dot{x}(t)\approx \frac{\omega_{0}a}{2\pi}2\sinh \{\frac{F_{ext}a}{2\sigma}\}
\exp \{-\frac{\Delta U_{0}}{\sigma}\}. 
\end{equation}
In the case that $F_{ext}a\gg \sigma$, the above equation can be approximated 
as 
\begin{equation}
\label{f_ext}
F_{ext}=\frac{2\Delta U_{0}(\phi)}{a}+\frac{2\sigma}{a}\ln (\frac{2\pi \dot{x}}
{\omega_{0}a}). 
\end{equation}
We now proceed to derive $U_0(\phi)$ by considering the case of
constant velocity motion ($\dot{x}=V$). In this case $\phi=D_0/V$, and
the friction force (given in Eq.~\ref{mu_dV}) is equal to the external force $F_{ext}$ 
given in Eq.~\ref{f_ext}. This gives rise to 
$$Mg(a_v-b_v\ln(V/V_0)) = 
\frac{2\Delta U_{0}(\phi)}{a}+\frac{2\sigma}{a}\ln (\frac{2\pi V}{\omega_0 a}).$$
By using the average contact age $\phi$, we can rewrite the above equation as
\begin{equation}
\Delta U_{0}(\phi)=\frac{a}{2}Mg\{a_{v}-b_{v}\ln \frac{D_{0}}{V_{0}}-A\ln 
\frac{2\pi D_{0}}{\omega_{0}a}+(b_{v}+A)\ln \phi \} ,
\end{equation}
where $A=2\sigma /aMg$. Given $U_0(\phi)$, we can write
the friction coefficient (which is equal to the external force divided by the
weight $Mg$ under the quasi-stationary approximation) in term of $\phi$ 
\begin{equation} 
\label{mu_phi}
\mu(\phi ,\dot{x})  =  a_{v}+b_{v}\ln \frac{\phi V_{0}}{D_{0}}
+A\ln \frac{\phi \dot{x}}{D_{0}}.
\end{equation}
When the velocity is not a constant, the contact age $\phi$ is assumed to
satisfy the following equation 
\begin{equation} 
\label{dot_phi}
\dot{\phi}(t)  =  1-\frac{\dot{x}\phi}{D_{0}}.  
\end{equation} 
By using this equation, the rate-dependence of the dynamic friction and the
contact time dependence of the static friction can be taken into account properly.
Eqs.\ref{mu_phi} and \ref{dot_phi} form the basis for the phenomenological description of
dry friction given in Ref. \cite{heslot}.

We now show that this description of friction is in fact equivalent to 
the state- and rate-dependent friction law proposed by Rice and Ruina \cite{ruina}
in the case of small deviation from constant velocity motion.
To make this connection, we choose $\theta=\ln (\phi V_{0}/D_{0})$ ($\theta$ is
the state variable in the Rice-Ruina theory), then Eqs.\ref{mu_phi} and \ref{dot_phi} become 
\begin{eqnarray}
\label{mu_theta1}
\mu(\theta ,\dot{x}) & = & a_{v}+A\ln \frac{\dot{x}}{V_{0}}+(b_{v}+A)
\theta , \\ 
\label{dot_theta1}
\dot{\theta}(t) & = & \exp \{ -\theta +\ln \frac{V_{0}}{D_{0}} \} 
-\frac{\dot{x}}
{D_{0}}. 
\end{eqnarray} 
In Rice-Ruina theory, the dynamical equations for the friction coefficient $\mu$
and state variable $\theta$ in the state- and rate-dependent friction law \cite{ruina} 
can be expressed as \cite{rice} 
\begin{eqnarray}
\label{mu_theta2}
\mu(\theta,\dot{x}) & = & \mu_{0}+A'\ln \frac{\dot{x}}{V_{0}}+B'\theta ,\\ 
\label{dot_theta2}
\dot{\theta}(t) & = & -\frac{\dot{x}}{D_{0}}\{ \theta +\ln \frac{\dot{x}}
{V_{0}} \}.
\end{eqnarray}
Consider the case that there is only a small deviation from the steady sliding state, we
write 
\begin{eqnarray*}
\dot{x}(t) & = & V+\Delta \dot{x}, \\ 
\theta (t) & = & -\ln \frac{V}{V_{0}}+\Delta \theta. 
\end{eqnarray*} 
Expending Eqs.~\ref{dot_theta1} and \ref{dot_theta2}
 to first order of $\Delta \dot{x}$ and $\Delta 
\theta $ give the same dynamical equation for $\Delta \theta $, 
\begin{equation} 
\frac{d(\Delta \theta )}{dt}=-\frac{V\Delta \theta}{D_{0}}-\frac{\Delta \dot
{x}}{D_{0}}. 
\end{equation} 
By choosing $a_{v}=\mu_{0}$, $A=A'$ and $b_{v}+A=B'$, Eqs.~\ref{mu_theta1} and 
\ref{mu_theta2} are exactly the same. Thus these two descriptions are in fact equivalent in the case where the deviation from the steady sliding motion is small. This is a strong indication that the 
phenomenological theory of dry friction discussed
in Ref.~\cite{heslot} is not limited to the explanation of experimental results on the paper-on-paper system, but a rather general theory that 
can be used as a basis for studying 
the dynamics involving dry friction on a range of materials. For example, it will be useful for studying rock friction, which is important in the 
study of earthquake dynamics.

For the case where $\sigma$ is comparable with $F_{ext}a$ (this can be thought 
of as in the \emph{static friction} regime), we have to use the original 
equation, i.e.\ Eq.\ 8, which can be rewritten as 
\begin{equation}
F_{ext}(\phi ,\dot{x})=MgA\sinh^{-1}\{ \frac{\pi \dot{x}}{\omega_{0}a}
\exp [\frac{\Delta U_{0}(\phi)}{\sigma}] \}.  
\end{equation}
The friction coefficient is then given as (with quasi-stationary approximation)
\begin{equation}
\label{mu_phi_new}
\mu (\phi ,\dot{x})=Asign(\dot{x})\sinh^{-1}\{ \frac{1}{2}\exp [\frac{\bar{
\mu}}{A}] \}, 
\end{equation}
where 
\begin{equation}
\label{bar_mu}
\bar{\mu}=a_{v}+b_{v}\ln \frac{\phi V_{0}}{D_{0}}+A\ln \frac{\phi |\dot{x}|}
{D_{0}}; 
\end{equation}
 here, $\bar{\mu}$ is an approximation of $\mu $ we describe in Eq.~\ref{mu_phi}. In this 
\emph{static friction} regime, the velocity $\dot{x}$ is very small and it can 
also change its sign, thus the equation for $\phi $ needs to be modified. It 
is reasonable to simply use 
\begin{equation}
\label{dot_phi_new}
\dot{\phi}(t)=1-\frac{|\dot{x}|\phi}{D_{0}}.  
\end{equation}

There are a few advantages of using Eqs.\ref{mu_phi_new}, \ref{bar_mu}, and \ref{dot_phi_new} to describe 
friction force: ($i$) there is no need to make a distinction between \emph{static} and \emph{dynamic} frictions, thus there is no \emph{stopping} 
condition (which is used to indicate the crossover from the ``dynamic'' friction regime
to the ``static'' friction regime) to worry about; ($ii$) the contact age is 
a well-defined quantity with Eq.~\ref{dot_phi_new}; ($iii$) There is no numerical singularity 
in the dynamical equations for the sliding block when the velocity goes to zero and changes sign. 
 
To include the inertial regime, we have to take into account the velocity 
strengthening described by Eq.~\ref{mu_d_highV}; we include this effect by simply adding a damping 
term $\eta \dot{x}$ in Eq.~\ref{mu_phi_new}. 

\section{Phase Diagram of Dry Friction Dynamics}
 
We now present the numerical results from the simulation of the block-spring system. Using the friction law described in the previous section, 
we can write down the dynamical equations of the block-spring system 
shown in Fig.~1. 
\begin{equation} 
M\ddot{x}(t)=K(Vt-x)-\mu (\phi ,\dot{x})N, 
\end{equation} 
where 
\begin{equation}
\label{mu_complete}
\mu (\phi ,\dot{x})=Asign(\dot{x})\sinh^{-1}\{ \frac{1}{2} \exp 
[\frac{\bar{\mu}}{A}] \}+\eta \dot{x}, 
\end{equation}
\begin{equation}
\bar{\mu}=a_{v}+b_{v}\ln \frac{\phi V_{0}}{D_{0}}+A\ln 
\frac{\phi |\dot{x}|}{D_{0}},
\end{equation}
and 
\begin{equation}
\dot{\phi}(t)=1-\frac{|\dot{x}|\phi }{D_{0}}.  
\end{equation}

We compare our result for the friction coefficient $\mu=\mu(D_0/V, V)$ with the experimental results when the sliding motion is steady ($\dot{x}=V$ 
and $\phi =D_{0}/V$). All the coefficients in the friction law 
are determined experimentally for the paper-on-paper system except for 
the value of $\eta$, which is determined
as follows. We consider the deflection point at $V=V^*$
from velocity weakening to velocity strengthening; this is determined using
$d\mu (D_{0}/V,V)/dV\mid_{V=V^{*}}=0$. This leads to 
\begin{equation}
\label{eta_eqn}
\eta =\frac{b_{v}}{2V^{*}}\frac{\exp [\frac{\bar{\mu}}{A}]}{\sqrt{\frac{1}{4}
\exp [\frac{2\bar{\mu}}{A}]+1}}, 
\end{equation}
where $\bar{\mu}=a_{v}-b_{v}\ln (V^*/V_{0})$. Substituting the experimental values
of $a_{v}=0.369$, $b_{v}=0.014$, $V_{0}=1\mu m/s$, $V^{*}=1mm/s$ \cite{heslot}
 and $A=0.011$ \cite{baumberger2} into Eq.~\ref{eta_eqn} gives $\eta=14.0s/m$.  

Using the values given above, the friction coefficient $\mu (D_{0}/V,V)$ in 
steady sliding region is shown in Fig.~2, which shows an excellent 
agreement with experiments \cite{heslot}. 
The logarithmic velocity weakening behavior of $\mu (D_{0}/V,V)$ can be found at low velocities ($\leq 0.1mm/s$); in this regime, the motion 
of the slider is creep-dominated and the effect of the damping 
term $\eta V$ is negligible. 
In the high-velocity regime, $\mu (D_{0}/V,V)$ becomes an increasing 
function of $V$ which behaves as $\eta V$ asymptotically. Hence, the 
experimentally observed dynamical friction coefficient $\mu_{d}(V)$ \cite{heslot} can 
be described by our simple expression of $\mu (D_{0}/V,V)$ without imposing 
any extra condition. 

We explore the dynamical phase diagram 
in control parameter space $(V,K/M)$ by systematically varying $V$ (within 
$10^{-2}$ to $10^{5}\mu m/s$) and $K$ (within $10^{-1}$ to $10^{5}N/cm$) with $M=1.2kg$ (this is the value used in the experiment). The phase 
diagram is shown in Fig.~3. Stick-slip occurs below a
 bifurcation curve $(K/M)_{c}(V)$, and steady sliding is found above (or to 
right of) it. The dots refer to the first steady sliding motions observed 
when increasing $K$ or $V$. The phase diagram consists of a creep-dominated 
regime which can be characterized by the time scale $\tau
_{cr}=D_{0}/V$ at low velocities and an inertial regime with a 
characteristic time $\tau_{in}=2\pi (M/K)^{1/2}$ at higher velocities. 
The dashed line indicates $\tau_{cr}=\tau_{in}$. In the creep-dominated 
(low-velocity) regime, 
the bifurcation occurs at a constant $K/M$ ($\approx 1.58\times 10^
{3}N/cm\cdot kg$). Fig.~4(a) shows the bifurcation from stick-slip to steady 
sliding as $K$ is increased in this regime. The bifurcation in the 
inertial (high-velocity) regime, on the other hand, occurs at a 
constant $V$ ($\approx 1.28\times 10^{3}\mu m/s$). The 
bifurcation from stick-slip to steady sliding in this 
regime is shown in Fig.~4(b). Except for the finite slope of the 
experimental bifurcation curve which we are not able to reproduced, 
our theoretical phase diagram agrees reasonably well with the 
experimental phase diagram. 

To investigate the nature of the bifurcation, we have measured the amplitude 
and period of the stick-slip oscillation by using different $K$ and $V$. The 
results are shown in Fig.~5. The amplitude and period decrease as 
the stiffness $K$ is increased. 
The amplitude approaches zero continuously as $K$ becomes larger and larger; 
this suggests that the bifurcation from stick-slip to steady sliding by 
increasing $K$ is a direct Hopf bifurcation. This agrees with previous
theoretical and numerical analysis \cite{heslot,baumberger2}.
On the other hand, the transition from stick-slip with a finite amplitude to steady sliding (with zero amplitude) when increasing the pulling 
velocity $V$ is rather sharp (the transition velocity is around 
$1.28\times 10^{3}\mu m/s$). This supports that the bifurcation 
encountered by increasing velocity is of the inverted Hopf type as 
was suggested previously \cite{baumberger1,heslot,persson} based on
experimental data (but was not understood theoretically). We have shown
numerically that the inverted Hopf bifurcation can be 
obtained within our fricition model.  

\section{Conclusion} 

We have shown that the phenomenological description of dry friction
proposed by Baumberger et al.\ 
is equivalent to the friction law of Rice and Ruina in the case of
small deviation from the constant sliding motion. We extend
the phenomenological description of Baumberger et al.\ and construct 
a simple model that can be used to simulate dynamics of dry friction
in various regimes. We have shown that, by associating 
the thermally activated creeping motion with a damping term which is 
significant only when the velocity is large, the 
model gives rise to satisfactory 
agreement with the experimental phase diagram. Except for 
the finite slope of the bifurcation curve observed in experiments, the 
essential features of the bifurcation in both creep-dominated and inertial 
regime can be reproduced using our simple model. We believe that the model
will also be very useful for numerical study of earthquake models which are
often modeled using block-spring systems. 
\pagebreak

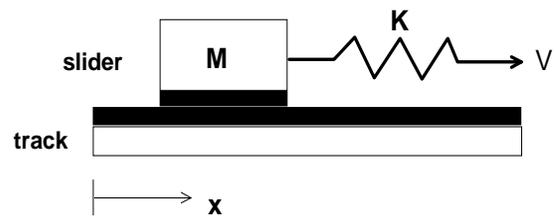
\begin{figure}
\caption{Schematic block-spring system. The slider with mass $M$ 
is driven by a pulling velocity $V$ through a spring with stiffness $K$. 
The displacement of the center of mass of the slider with respect to the 
track is $x$.}
\label{block_spring}
\end{figure}

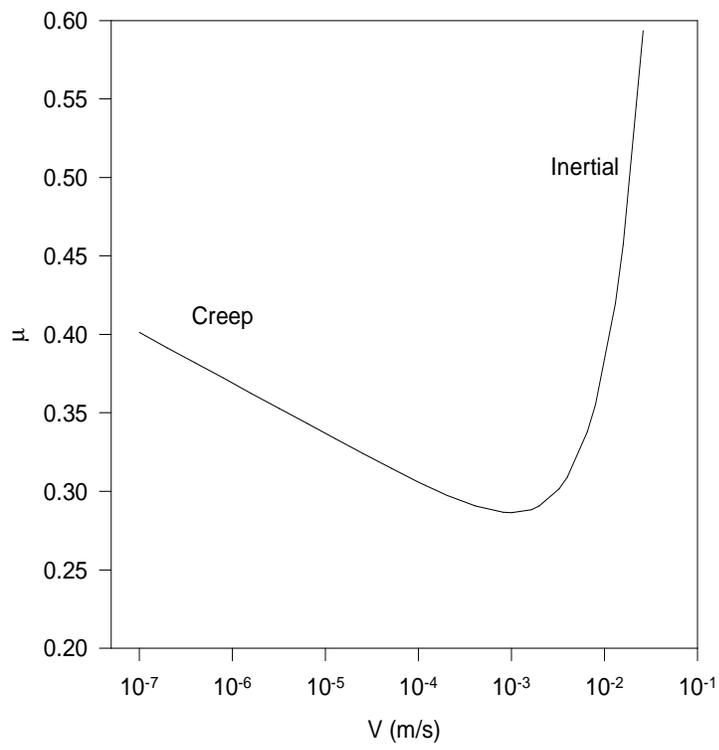
\begin{figure}
\caption{Friction coefficient $\mu $ described by Eq.~\ref{mu_complete} vs the 
pulling velocity $V$ when the sliding motion is steady. 
$\mu $ changes from velocity weakening to velocity strengthening at 
$V=V^{*}(=1mm/s)$; this gives $\eta =14.0s/m$. }
\label{mu_vs_V}
\end{figure}

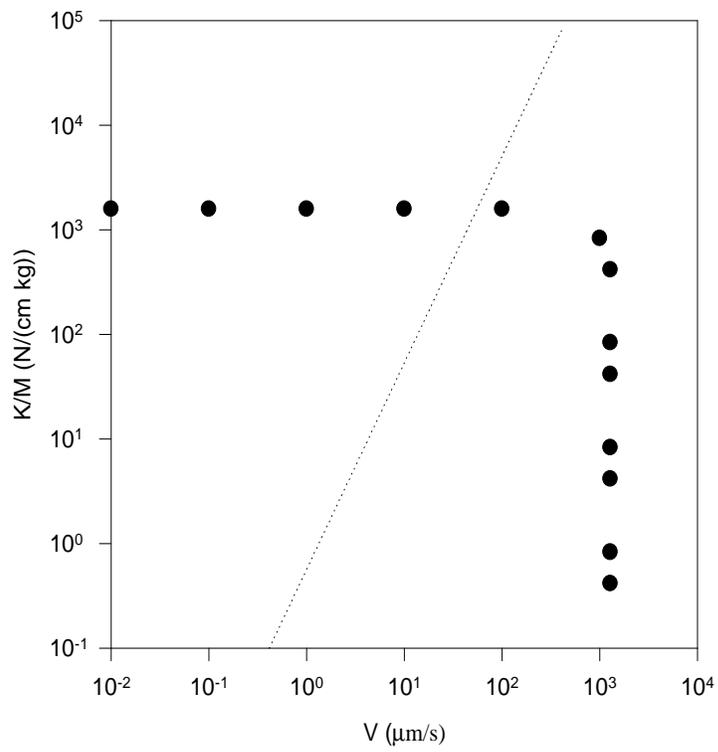
\begin{figure}
\caption{Dynamical phase diagram in $(V,K/M)$ space, 
obtained by varying 
$K$ and $V$ with $M=1.2kg$. It consists of a creep-dominated regime 
which can be characterized by the time scale $\tau_{cr}$ and an inertial 
regime with a characteristic time $\tau_{in}$ (see text). The dashed 
line indicates $\tau_{cr}=\tau_{in}$. The bifurcation curve which is 
represented by dots, changes from a horizontal line 
$(K/M\approx 1.58\times 10^{3} N/cm\cdot kg)$ to a vertical line 
$(V\approx 1.28\times 10^{3}\mu m/s)$ at the crossover from creep to 
inertial motion. }
\label{phase_diagram}
\end{figure}

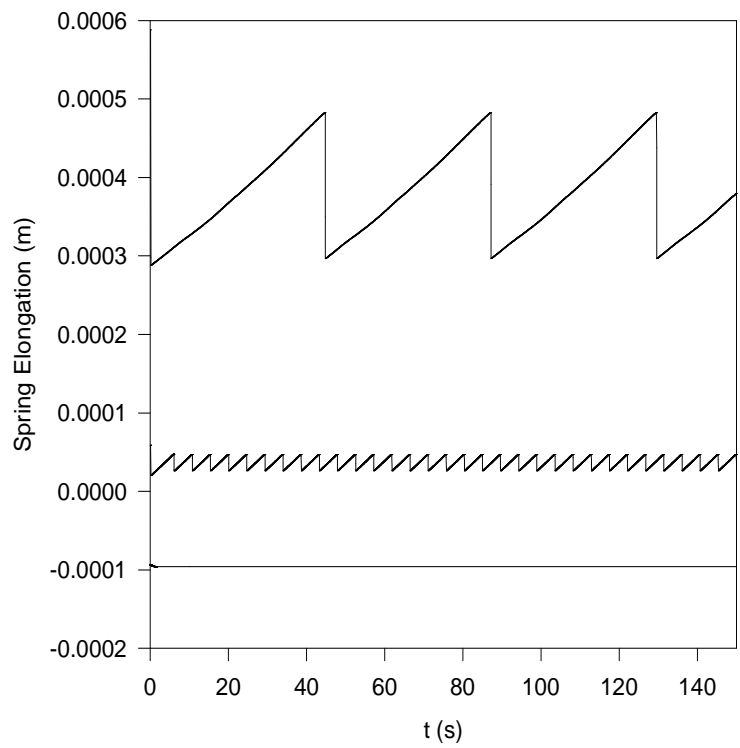
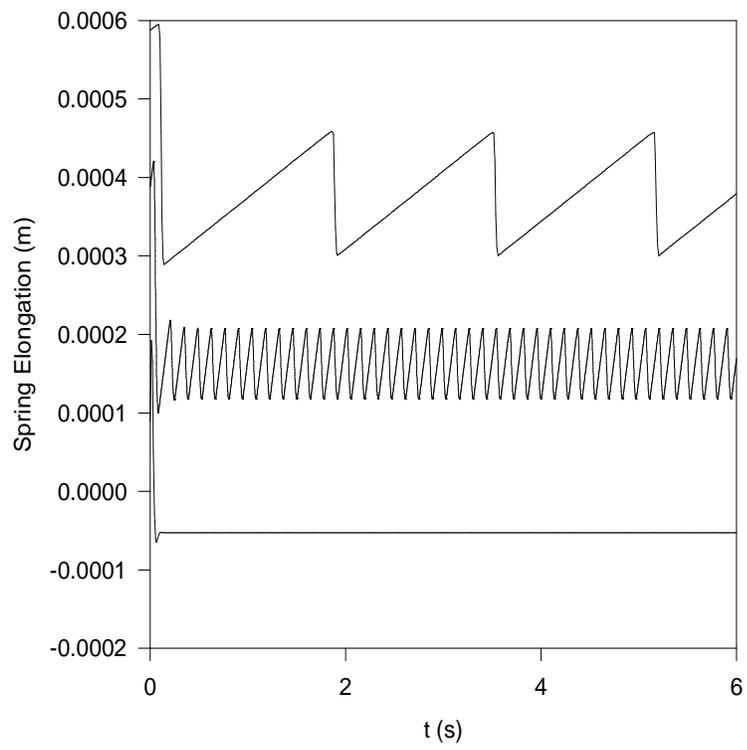
\begin{figure}
\caption{(a) Time evolution of the spring elongation, when crossing 
the bifurcation curve in low-velocity regime. $V=5.0\mu m/s$, $M=1.2Kg$, and 
(from upper to lower curve) $K=10^{2}, 10^{3}, 10^{4}N/cm$. The lowest curve 
has been shifted vertically by the amount $-1.0 \times 10^{-4}$ for the sake of clarity. 
(b) The transition from stick-slip to steady sliding in 
high-velocity regime. $K=10^{2}N/cm$, $M=1.2Kg$, and (from upper to lower 
curve) $V=10^{2}, 10^{3}, 10^{4}\mu m/s$. The lowest and the second lowest 
curves have been 
shifted vertically by the amoumts $-5.0\times 10^{-4}$ and $-2.0\times 10^{-4}$
respectively for the sake of clarity.}
\label{spring_elongation}
\end{figure}

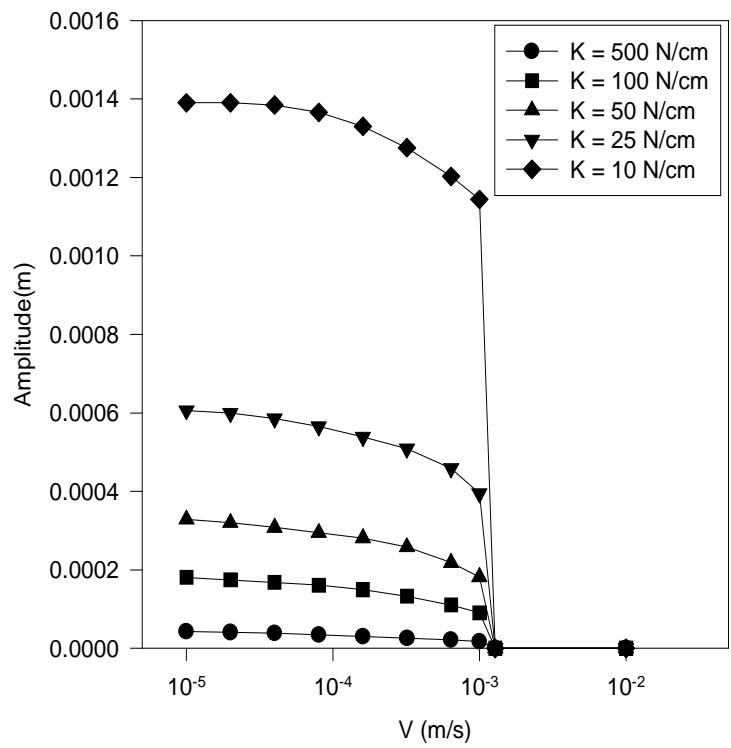
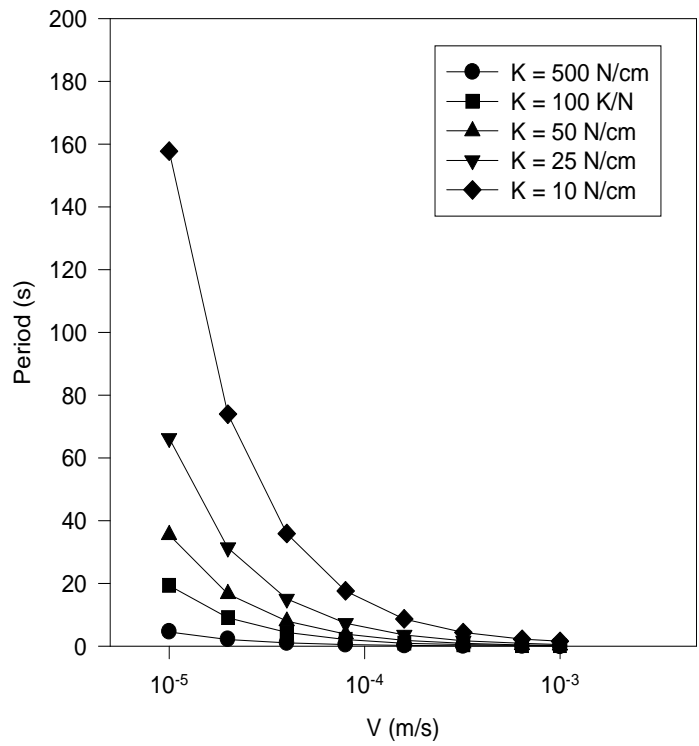
\begin{figure}
\caption{(a) Amplitude of the stick-slip oscillations vs the pulling velocity
$V$ at various values of $K$. The amplitude approaches zero continuously 
as $K$ increases, whereas the transition from stick-slip to steady sliding 
as $V$ is increased is rather sharp.(b) Period of the stick-slip oscillations vs 
the pulling velocity $V$ at various $K$. The period approaches zero continuously 
either as $K$ or $V$ increases. }
\label{amplitude}
\end{figure}

\end{document}